\documentclass{article}

\usepackage{amssymb,amsfonts,amsmath}
\usepackage{cite,enumerate,float,indentfirst}
\usepackage{color}

\def\be{\begin{eqnarray}}
\def\ee{\end{eqnarray}}
\def\nn{\nonumber}

\def\p{\partial}

\def\Tr{{\rm Tr}\,}

\definecolor{red}{rgb}{1,0,0}
\definecolor{orange}{rgb}{1,0.5,0}
\definecolor{violet}{rgb}{0.7,0,1}

\hoffset= - 1.0in         

\begin{document}

\title{\vspace{.5cm}{\LARGE {\bf  Discrete Painlev\'e equation, Miwa variables\\
and string equation in 5d  matrix models}}
\author{
{\bf A. Mironov$^{a,b,c}$}\footnote{mironov@lpi.ru; mironov@itep.ru},
\ {\bf A. Morozov$^{d,b,c}$}\thanks{morozov@itep.ru}\ \ and
{\bf Z. Zakirova$^{e}$}\footnote{zolya\_zakirova@mail.ru}
}
\date{ }
}

\maketitle

\vspace{-6.3cm}

\begin{center}
\hfill FIAN/TD-16/19\\
\hfill IITP/TH-23/19\\
\hfill ITEP/TH-15/19\\
\hfill MIPT/TH-13/19
\end{center}

\vspace{3cm}

\begin{center}
$^a$ {\small {\it Lebedev Physics Institute, Moscow 119991, Russia}}\\
$^b$ {\small {\it ITEP, Moscow 117218, Russia}}\\
$^c$ {\small {\it Institute for Information Transmission Problems, Moscow 127994, Russia}}\\
$^d$ {\small {\it MIPT, Dolgoprudny, 141701, Russia}}\\
$^e$ {\small {\it Kazan State Power Engineering University, Kazan, Russia}}
\end{center}

\vspace{.5cm}

\begin{abstract}
The modern version of conformal matrix model (CMM) describes conformal blocks
in the Dijkgraaf-Vafa phase.
Therefore it possesses a determinant representation and becomes a Toda chain
$\tau$-function only after a peculiar Fourier transform in internal dimensions.
Moreover, in CMM Hirota equations arise in a peculiar discrete form
(when the couplings of CMM are actually Miwa time-variables).
Instead, this integrability property is actually independent of the measure
in the original hypergeometric integral.
To get hypergeometric functions, one needs to pick up a very special $\tau$-function,
satisfying an additional ``string equation".
Usually its role is played by the lowest $L_{-1}$ Virasoro constraint,
but, in the Miwa variables, it turns into a finite-difference equation with respect to
the Miwa variables.
One can get rid of these differences by rewriting the string equation
in terms of some double ratios of the shifted $\tau$-functions,
and then these ratios satisfy more sophisticated equations
equivalent to the discrete Painlev\'e equations by M. Jimbo and H. Sakai ($q$-PVI equation).
They look much simpler in the $q$-deformed ($"5d"$) matrix model,
while in the "continuous" limit $q\longrightarrow 1$ to $4d$
one should consider the Miwa variables with non-unit multiplicities,
what finally converts the simple discrete Painlev\'e $q$-PVI into sophisticated
differential Painlev\'e VI equations, which will be considered elsewhere.
\end{abstract}

\section{Introduction}

The purpose of this paper is to demonstrate
that the discrete Painlev\'e equations $q$-PVI \cite{JS}
play the role of string equation in ``$5d$" conformal \cite{confmamo}
matrix model (CMM), which underlies \cite{MMS} the theory of AGT-related \cite{AGT}
conformal blocks \cite{CFT} and Nekrasov functions \cite{Nek}.
As all matrix models \cite{UFN3},
the CMM satisfies a set of Ward identities (Virasoro constraints) \cite{vircon},
which, in this particular case, {\bf can be reduced to a small moduli space}
of just a few $\alpha$-couplings.
Also, {\bf after a peculiar Fourier transform in the matrix size $N$, it possesses
a determinant representation} \cite{MMZ} and therefore is a $\tau$-function
of integrable hierarchy, i.e. satisfies discrete Hirota equations in Miwa variables.
As usual, an interesting point is the interplay between integrability
and Virasoro constraints.
The basic difference is that integrability does not depend on the integration measure
of matrix model
and, in this sense, is a pure classical property independent of quantization
of the theory.
At the same time, the Ward identities strongly depend on the measure and remain
independent only of the choice of integration contour.
This means that the Virasoro constraints are picking up a peculiar narrow class
of $\tau$-functions,
they are naturally named ``matrix-model $\tau$-functions", and finding a nice
way to describe this class is one of the central problems in non-perturbative physics.
Usually, when applied to $\tau$-functions, the Virasoro constraints reduce to just
a single independent ``string equation",
and, in this sense, the problem is to understand what the string equations can be.
This explains a significance of the claim \cite{MMpain12}
that the Painlev\'e VI equation, whose solutions were associated with conformal blocks in \cite{Iorg},
can in fact be exactly the string equation for the CMM.
A natural question is why it is so complicated, why just the usual $L_{-1}$ constraint
is not the right answer?
The reason is that the Virasoro constraints include $\tau$-functions at different
(shifted) values of the Miwa variables  $\alpha$,
and the Painlev\'e equation emerges when one rewrites them
in another form, without such shifts.
This is achieved by switching to some special combinations $w_1$ and $w_2$ of
shifted $\tau$, and these two $w$'s appear related by a pair of equations,
equivalent to the Painlev\'e VI equation
(in the simplest case of the {\it 4-point} conformal block).
The point here is that the CMM actually has {\it two} sets of parameters:
in addition to the $\alpha$-couplings, there are "background" points $z$,
and the Painleve\'e equation is a differential equation with respect to $z$
(which is a double ratio of the four punctures: positions of vertex
operators in the conformal block).
Moreover, it can be naturally split into an algebraic relation between $w_1$ and $w_2$,
which is just a Seiberg-Witten spectral curve, and a true differential equation,
which describes its $z$-dependence.
Actually, all these properties look much simpler in the  $q$-deformed (``$5d$") CMM,
because there the two discreteness, the built-in one in the couplings $\alpha$
and the $q$-related one in the background parameter $z$ become nearly undistinguishable
(this property is sometimes called {\it duality} between the Coulomb and Higgs branches,
which gets transparent after the lift from $4d$ to $5d$).
Therefore, this letter will concentrate on the $q$-deformed CMM and discrete Painlev\'e equations $q$-PVI (see \cite{BGT} for a different relation of $q$-Painlev\'e with matrix models, see also \cite{GG}).

We begin in sec.2 from reminding the standard facts about usual matrix models and the discrete Painlev\'e equations $q$-PVI.
Then in sec.3 we remind the basic facts about Miwa variables in integrable systems and relation to CMM.
As already mentioned, the relation includes the Fourier transform of the naive CMM in $N$
and leads to determinant representation, which emerges after such a transform is performed.
After this, in sec.4 we discuss the $5d$ CMM in detail and demonstrate that its partition function solves in this case the discrete Painlev\'e $q$-PVI equations.
At last, in sec.5 we discuss the 8-equation system of \cite{Sakai},
the discrete Painlev\'e equations being derived from these 8 equations:
actually exactly one half of these eight are Hirota equations, i.e.
possess the property of measure-independence.
The other four equations are actually not Hirota ones, since they depend on the concrete hypergeometric solution to the Toda chain hierarchy, and therefore should be
equivalent to the string equation.
We devote a special sec.5.3  to demonstrate what is the meaning of these 8 equations and how
the Painlev\'e equation arises from them in the simplest case of $N=1$.

What we {\it do not do} in this paper, we do not actually {\it derive} the 8 equations
of \cite{Sakai} from CMM, we just confirm {\it an observation} that they are true. In fact,
as it was already mentioned, only four of them require a derivation, since the other four are just bilinear Hirota equations that follow from the fact that the matrix model partition function is a $\tau$-function of the Toda chain hierarchy. The derivation of remaining four equations is completely analogous to the derivation of \cite{Sakai}, and we do not repeat it here.
We are also rather brief on the story of Fourier transform and determinant
representations just referring to the original paper \cite{MMZ} for further details.
Finally, we do not perform the reduction from $5d$ to $4d$, where the simply looking
discrete Painlev\'e $q$-PVI equations turn into a sophisticated differential Painlev\'e VI equation.
All this is postponed to a big technical version of the present text.
The goal of this letter is just to make
the very claims and make them  precise, well grounded and justified.

\section{Matrix models and Painlev\'e equations}

In this section, we briefly describe a set of standard facts about matrix models and Painlev\'e equation necessary for the main body of the text.

\subsection{Matrix models and string equations}

The first issue is matrix models. We use the term ``matrix models'' for
arbitrary eigenvalue integrals with the Vandermonde-like factor in the measure,
though their matrix integral representations are not always that much simple.
Matrix models possess a set of defining properties \cite{UFN3,DVrev,more}:
\begin{itemize}
\item {\bf Ward identities.} The partition function of matrix model satisfies an infinite set of Ward identities.
\item {\bf Solutions.} The number of solutions to the Ward identities are parameterized by the number of independent closed contours in the eigenvalue integral representation of matrix model (when the solution is not unique, the model is said to be in the {\bf Dijkgraaf-Vafa phase} \cite{DV}).
\item {\bf Integrability.} The partition function of matrix model is related to a $\tau$-function of an integrable hierarchy: it is either the partition function or its Fourier transform (in the Dijkgraaf-Vafa phase \cite{MMZ}) which is the $\tau$-function.
\item {\bf String equation.} The concrete solution of the integrable hierarchy is fixed by the string equation(s), which is typically the lowest Ward identity(ies). Moreover, the full set of Ward identities is equivalent to the integrable hierarchy with only the string equation added.
\item {\bf Measure (in)dependence.} The measure in the eigenvalue integral is essentially
the Vandermonde-like factor responsible for a universal ``interaction" between the eigenvalues
times a product of additional {\it measure functions} for all eigenvalues.
Integrability properties do not depend on the choice of this measure function, only on the
Vandermonde.
Only the string equation is fully sensitive to the choice of the measure,
and this makes it so important
to specify the partition function of a particular matrix model within the relatively wide space of various $\tau$-functions.
\end{itemize}

\subsubsection{Hermitian Matrix model}

We start with the most simple and typical example of the matrix model: the Gaussian Hermitian matrix model. Ii is given by the following integral over $N\times N$ Hermitian matrices
\be\label{HMM}
Z_N\{t\} ={1\over \hbox{Vol}_{U(N)}}\int_{N\times N} dM \exp \left(-\frac{\eta}{2}\,\Tr M^2
+ \sum_k t_k \,\Tr M^k\right)
\ee
where $dM$ is the invariant measure on Hermitian $N\times N$ matrices and Vol$_{U(N)}$ is the volume of the unitary group $U(N)$.
\begin{itemize}
\item This partition function satisfies an infinite set of Ward identities, which form (a Borel subalgebra of) the Virasoro algebra:
\be
\hat L_nZ_N\{t\}:=\left(-\eta\frac{\partial}{\partial t_{n+2}}
+\sum kt_k \frac{\partial}{\partial t_{k+n}}
+ \sum_{a=1}^{n-1} \frac{\partial^2}{\partial t_a\partial t_{n-a}}
+2N\frac{\partial}{\partial t_n}
+ N^2\delta_{n,0}
\right)  Z_N\{t\}=0,
\ \ \ \ n\geq -1
\label{VirH}
\ee
\item It can be reduced to the eigenvalue integral
\be\label{Zt}
Z_N(t_k):={1\over N!}\int\prod_i dx_i\Delta^2(x)\exp\Big(-\sum_i\eta x_i^2+\sum_{k,i} t_kx_i^k\Big)
\ee
where $\Delta(x)$ is the Vandermonde determinant.
This integral is considered as a formal power series in time variables $t_k$ and, hence, is given just by moments of the Gaussian integral. Therefore, there is only one integration contour, which is the real axis, and only one solution to the Ward identities.
\item Integral (\ref{Zt}) can be rewritten as a determinant
\be\label{detrept}
Z_N(t_k)=\det_{i,j=1..N}C_{i+j-2},\ \ \ \ \ \ \ C_k:=\int_{\mathbb{R}}dxx^k\exp\Big(-\eta x^2+\sum_{m} t_mx^m\Big)
\ee
This determinant is nothing but a $\tau$-function of integrable Toda chain hierarchy.
\item One can consider instead of (\ref{Zt}) a more general eigenvalue integral
\be\label{Z4t}
Z_N(t_k):={1\over N!}\int\prod_i dx_i\mu(x_i)\Delta^2(x)\exp\Big(\sum_{k,i} t_kx_i^k\Big)
\ee
with an arbitrary measure function $\mu(x)$, then, in (\ref{detrept}), $C_k=\int_{\mathbb{R}}dx\mu(x)x^k\exp\Big(-\eta x^2+\sum_{m} t_mx^m\Big)$. This more general integral is still a $\tau$-function of integrable Toda chain hierarchy. The concrete solution (\ref{Zt}) is unambiguously picked up by the string equation additional to the integrable hierarchy
\be
\hat L_{-1}Z_N\{t\}=\left(-\eta\frac{\partial}{\partial t_{1}}
+\sum kt_k \frac{\partial}{\partial t_{k-1}}
\right)  Z_N\{t\}=0
\ee
\end{itemize}

\subsubsection{Kontsevich model\label{Kont}}

Another example is a matrix model that depends on the external matrix, Kontsevich model:
\be
\label{Kint}Z_K= {\int DX\ \exp\left(-{1\over
3}\hbox{Tr}X^3-\hbox{Tr}AX^2\right)\over \int DX\
\exp\left(-\hbox{Tr}AX^2\right)}\ee
which is a function of time-variables
\be
t_{2k+1}:=\displaystyle{{1\over 2k+1}\hbox{Tr} A^{-2k-1}}-{2\over
3}\delta_{k,3}
\ee
\begin{itemize}
\item The Kontsevich integral satisfies an infinite set of Virasoro constraints:
\be\label{K}
\hat L_{n}Z_K=\left( \sum_{k>0} \left(k+{1\over
2}\right)t_{2k+1}\frac{\p}{\p t_{2k+1+2n}} +
{1\over 4}\sum_{a+b=n-1}\frac{\p^2}{\p t_{2a+1}\p t_{2b+1}}+
\frac{\delta_{n,0}}{16}+\frac{\delta_{n,-1}t_1^2}{4}
\right)Z_K=0
\ee
\item The Kontsevich integral is understood as a formal power series in variables $t_k$, which fixes just a unique solution to the Virasoro constrains \cite{AMMP}.
\item $Z_K$ is a $\tau$-function of the KdV hierarchy \cite{GKM}, which is reduction from the KP $\tau$-function, which depends only on  the odd time variables $t_{2k+1}$.
\item The concrete solution to the KdV hierarchy is again unambiguously picked up by the first Virasoro constraint $\hat L_{-1}$, which is the string equation:
\be\label{stre}
\hat L_{-1}Z_K=\left( \sum_{k>0} \left(k+{1\over
2}\right)t_{2k+1}\frac{\p}{\p t_{2k-1}}+\frac{t_1^2}{4}
\right)Z_K=0
\ee
One can now leave only two non-zero time variables $t_1$ and $t_{3}$ and differentiate (\ref{stre}) w.r.t. $t_1$ in order to get an equation for $u:={\p^2\log Z_k\over\p t_1^2}$
\be
3t_3u+t_1=0
\ee
Similarly, choosing non-zero $t_1$ and $t_5$, one obtains the equation (with $t_5$ chosen a proper constant) \cite{FKN1}
\be\label{PI}
{1\over 3}{\p^2 u\over\p t_1^2}-u^2+t_1=0
\ee
which is the Panlev\'e I equation, etc. This is the first example where we obtain the Painlev\'e equation as a corollary of a reduction of the string equation to few (two) non-zero times.
\end{itemize}

\subsection{Discrete Painlev\'e equation}

As we already noted, the case of discrete Painlev\'e $q$-PVI equations turns out to be much simpler than the case of standard Painlev\'e VI equation. It is an equation for two functions $w_1(z)$ and $w_2(z)$, and has the form \cite{JS} (in fact, there are many other discrete Painlev\'e equations, see \cite{NSY} for a review)
\be\label{PVI}
{w_1(z)w_1(qz)\over a_3a_4}={(w_2(qz)-b_1z)(w_2(qz)-b_2z)\over (w_2(qz)-b_3)(w_2(qz)-b_4)}\nn\\
{w_2(z)w_2(qz)\over b_3b_4}={(w_1(z)-a_1z)(w_1(z)-a_2z)\over (w_1(z)-a_3)(w_1(z)-a_4)}
\ee
where the constants $a_i$, $b_i$ satisfy the constraint
\be\label{cond}
{b_1b_2\over b_3b_4}=q{a_1a_2\over a_3a_4}
\ee
By rescalings $w_1(z)$, $w_2(z)$ and $z$, one can always remove three of these constants $a_i$, $b_i$ so that remaining four constants we can always parameterize with four parameters.

Note that the continuous limit of these discrete Painlev\'e $q$-PVI equations to the Painlev\'e VI equation is quite tricky: one expands the equation nearby the point
\be
a_i=b_i=q=1
\ee
so that
\be
w_1={w_2-z\over w_2-1},\ \ \ w_2={w_1-z\over w_1-1},\ \ \ \hbox{i.e.}\ \ \ {(w_1-a_1z)(w_1-a_2z)\over (w_1-z)(w_1-1)}{1\over qw_2}=1
\ee
Now choosing
\be
q=1-\epsilon,\ \ \ a_i=1+\epsilon\mathfrak{a}_i,\ \ \ b_i=1+\epsilon\mathfrak{b}_i,\ \ \ y_1=w_1,\ \ \
{(w_1-a_1z)(w_1-a_2z)\over (w_1-z)(w_1-1)}{1\over qw_2}=1-\epsilon w_1y_2
\ee
with $\epsilon\to 0$, one arrives at a pair of first order differential equations for $y_1$, $y_2$ that are equivalent to the Painlev\'e VI equation \cite{JS}.

This is quite a surprise that such a fancy limit may naturally emerge, however, it turns out to be the case: it naturally emerges as the $4d$ limit of the $5d$ matrix model, which is nothing but the matrix integral with 3 arbitrary non-vanishing Miwa variables, or just a matrix model with a 3-logarithm potential (see sec.\ref{confmamo}).

\section{Matrix models in Miwa variables}

\subsection{Miwa variables and Hirota bilinear identities}

Now let us consider the change of variables $t_k$ in the integral (\ref{Z4t}) with an arbitrary measure function $\mu(x)$ to the so called Miwa variables $(z_a,2\alpha_a)$
\be
t_k:={1\over k}\sum_a 2\alpha_az_a^{-k}
\ee
with arbitrary many parameters $z_a$ and $\alpha_a$. Then, the integral becomes
\be\label{Z4M}
Z_N(z_a;\alpha_a):={1\over N!}\int\prod_i dx_i\mu(x_i)\Delta^2(x)\prod_{i,a} \Big(1-{x_i\over z_a}\Big)^{2\alpha_a}
\ee
As soon as (\ref{Z4t}) is a $\tau$-function of the Toda chain hierarchy \cite{Toda}, it satisfies the Hirota bilinear identities, which, in the Miwa variables, look like \cite{Miwa,GKM,versus,Kharchev}
\be\label{Hirota1}
(z_a-z_b)\cdot Z_N(\alpha_c+1/2)\cdot Z_N(\alpha_a+1/2,\alpha_b+1/2)+
(z_b-z_c)\cdot Z_N(\alpha_a+1/2)\cdot Z_N(\alpha_b+1/2,\alpha_c+1/2)+\nn\\
+(z_c-z_a)\cdot Z_N(\alpha_b+1/2)\cdot Z_N(\alpha_a+1/2,\alpha_c+1/2)=0
\ee
and are satisfied for all triples of $z_{a,b,c}$ and $\alpha_{a,b,c}$. It can be also derived from the determinant representations.

Similarly, for all pairs of $z_{a,b}$ and $\alpha_{a,b}$, there is another equation \cite{AZ}
\be\label{Hirota2}
(z_a-z_b)\cdot Z_N\cdot Z_{N-1}(\alpha_a+1/2,\alpha_b+1/2)-z_a\cdot Z_N(\alpha_a+1/2)\cdot Z_{N-1}(\alpha_b+1/2)+\nn\\
+z_b\cdot Z_N(\alpha_b+1/2)\cdot Z_{N-1}(\alpha_a+1/2)=0
\ee
and
\be\label{Hirota22}
z_b\cdot Z_N\cdot Z_{N-1}(\alpha_a+1/2,\alpha_b+1/2)- Z_N(\alpha_a+1/2)\cdot Z_{N-1}(\alpha_b+1/2)-\nn\\
-z_b\cdot Z_N(\alpha_b+1/2)\cdot Z_{N-1}(\alpha_a+1/2)=0
\ee
if $z_a=0$.
In fact, it follows from (\ref{Hirota1}), since changing multiplicity by one unit $\alpha\to\alpha+1/2$ is equivalent to inserting a fermion in the fermionic realization of the Toda hierarchy \cite{JimboMiwa,versus,Kharchev}, and such is increasing the discrete Toda time $N$ by one: $Z_N\to Z_{N+1}$ as well. Similar identities that involve three multiplicities are
\be\label{Hirota3}
z_b\cdot Z_N(\alpha_c-1/2)\cdot Z_{N-1}(\alpha_a+1/2,\alpha_b+1/2)- Z_N(\alpha_a+1/2,\alpha_c-1/2)
\cdot Z_{N-1}(\alpha_b+1/2)-\nn\\
-z_b\cdot Z_N(\alpha_b+1/2,\alpha_c-1/2)\cdot Z_{N-1}(\alpha_a+1/2)=0
\ee
\be\label{Hirota4}
z_c\cdot Z_{N-1}\cdot Z_N(\alpha_a-1/2,\alpha_b-1/2,\alpha_c-1/2)- Z_{N-1}(\alpha_a-1/2)
\cdot Z_N(\alpha_b-1/2,\alpha_c-1/2)-\nn\\
-z_c\cdot Z_{N-1}(\alpha_c-1/2)\cdot Z_N(\alpha_a-1/2,\alpha_b-1/2)=0
\ee
if $z_a=0$. There is also a bilinear difference equation that relates $Z_{N+1}$ and $Z_{N-1}$ \cite{AZ}, but we do not need it here.

\subsection{Conformal matrix models and Painlev\'e VI equation\label{confmamo}}

Now let us note that the integrals of the form (\ref{Z4M}) naturally emerge in studying the Virasoro conformal blocks within the conformal matrix model approach \cite{confmamo,MMS}. Indeed, the CMM-representation of the standard Virasoro conformal block of the theory with central charge $c=1$ with conformal dimensions parameterized by conformal momenta, $\Delta_i=\alpha_i^2$ is given by the formula
\be
B^{(4d)}(\alpha_i;\alpha; z)=z^{\Delta-\Delta_1-\Delta_2 }\cdot\left( 1+ {(\Delta_2-\Delta_1+\Delta)(\Delta_3-\Delta_4+\Delta)\over 2\Delta}\cdot z+{\cal O}(z^2)\right)=\mathfrak{Z}^{(4d)} \cdot Z_{N_1,N_2}^{(4d)}
\ee
with the eigenvalue (matrix) model integral
\be\label{ZB4}
Z_{N_1,N_2}^{(4d)}=z^{2\alpha_1\alpha_2}(1-z)^{2\alpha_2\alpha_3}
\cdot{1\over N_1!N_2!}
\int\prod_i dx_i\Delta^2(x)\prod x_i^{2\alpha_1}(z-x_i)^{2\alpha_2}(1-x_i)^{2\alpha_3}
\ee
where $\mathfrak{Z}$ is a normalization factor, and the matrix integral (\ref{ZB4}) depends on two integers, $N_1$ and $N_2$ that count the number of integrations over the contours $C_1=[0,z]$ and $C_2=[1,\infty)$ respectively. These integers are determined by the external conformal momenta $\alpha_i$ and the internal one, $\alpha$:
\be\label{N}
N_1=\alpha-\alpha_1-\alpha_2,\ \ \ \ \ N_2=-\alpha-\alpha_3-\alpha_4
\ee
This is a typical Dijkgraaf-Vafa type model with two different contours, its partition function is not a $\tau$-function of an integrable hierarchy. In order to have a $\tau$-function, one can consider a model with the same measure and with the unique integration contour given by a formal sum of two contours $C(\mu_1,\mu_2):=\mu_1\cdot C_1+\mu_2\cdot C_2$, where $\mu_1$ and $\mu_2$ are formal parameters:
\be\label{MM}
Z_N^{(4d)}(\mu_1,\mu_2):={1\over N!}\int_{C(\mu_1,\mu_2)}\prod_i dx_i\Delta^2(x)\prod x_i^{2\alpha_1}(z-x_i)^{2\alpha_2}(1-x_i)^{2\alpha_3}
\ee
Then, we immediately have
\be
Z_N^{(4d)}(\mu_1,\mu_2)=\sum_{N_1,N_2:\ {N_1+N_2=N}}
\mu_1^{N_1}\mu_2^{N_2}\cdot Z_{N_1,N_2}^{(4d)}
\label{FT4}
\ee
i.e. $Z_N^{(4d)}(\mu_1,\mu_2)$ is a generation function of the Dijkgraaf-Vafa partition functions $Z_{N_1,N_2}^{(4d)}$. This is nothing but a discrete Fourier transform in the variable $\mu_1/\mu_2$ with the sum $N=N_1+N_2$ fixed.

$Z_N^{(4d)}(\mu_1,\mu_2)$ is already a $\tau$-function of the Toda chain in Miwa variables (\ref{Z4M}) restricted to the point with only three non-zero Miwa variables. As any matrix model $\tau$-function, the multiple integral (\ref{MM}) has the standard determinant representation (\ref{detrept})
\be
Z_N^{(4d)}(\mu_1,\mu_2)= z^{2\alpha_1\alpha_2}(1-z)^{2\alpha_2\alpha_3}
\cdot\det_{1\le i,j\le N}G(i+j-2)
\label{detrep4}
\ee
where
\be\label{G4}
G(k)= \mu_1 \int_0^z x^{2\alpha_1+k}(z-x)^{2\alpha_2}(1-x)^{2\alpha_3}dx
+ \mu_2\int_1^\infty x^{2\alpha_1+k}(z-x)^{2\alpha_2}(1-x)^{2\alpha_3}dx
\label{Gfun4}
\ee
and it was demonstrated in \cite{MMpain12} that it satisfies the Painlev\'e VI equation.

Thus, it is the set-up where the continuous limit of the discrete Painlev\'e equations (\ref{PVI}) naturally emerges. In the remaining part of the paper we demonstrate that the same scheme is equally well applicable to the $q$-Painlev\'e case of $5d$ conformal matrix models.
Moreover, the structures behind the Painlev\'e equation in this discrete $q$-case are much more transparent than in the $4d$ model.

\section{5d matrix model and discrete Painlev\'e equations}

\subsection{CMM representation of the $q$-Virasoro conformal block}

In the $q$-Virasoro case, the procedure is literally the same: at the first step, we realize the conformal block by the matrix integral \cite{5dDF,Ito,NZ}. There are only two differences with the Virasoro case: first, all integrals become the Jackson integrals, and, second, some powers are replaced with the Pochhammer symbols. The Jackson integral is defined as a sum
\be\label{Ji}
\int_0^1f(x)d_qx=(1-q)\sum_{k=0}^\infty f(q^k)
\ee
One can transform eigenvalue integrals over the contours $C_1=[0,z]$ and $C_2=[1,\infty)$ into the integrals over $C=[0,1]$ with the changes of variables: $x\to zu$ and $x\to 1/v$ respectively. These integrals can be immediately deformed to the Jackson integrals in form (\ref{Ji}). One has also to substitute the degrees $\alpha_2$ and $\alpha_3$ in (\ref{ZB4}) with the $q$-Pochhammer symbols\footnote{This is the case for the integer values of $p$, the extension to non-integer is immediate:
$$
(\xi;q)_p\to \exp\Big(-\sum_{k=1}^\infty {1-q^{pk}\over 1-q^k}{\xi^k \over k}\Big)
$$
}:
\be
(1-\xi)^p\to (\xi;q)_p=\prod_{k=0}^{p-1}(1-q^k\xi)
\ee
After making these two changes, we immediately arrive to the CMM representation
of the $q$-Virasoro conformal block of the theory with central charge $c=1$ (see \cite{BGM} for $c\ne 1$ case), the counterpart of (\ref{ZB4}), \cite{5dDF}:
\be
B^{(5d)}(\Delta_i;\Delta; z)=\mathfrak{Z}^{(5d)} \cdot Z_{N_1,N_2}^{(5d)}
\ee
with\footnote{Note that in the literature, the prefactor $(z;q)_{2\alpha_2\alpha_3}$
is often omitted (see, e.g., \cite[Eq.(4.29)-(4.30)]{Ito}).
This factor is due to the additional $U(1)$ group that participates in the AGT conjecture,
and would be necessary if one requires that the $q$-Virasoro conformal block turns into
the Virasoro one when $q\to 1$.
It is also present in the solution to the continuous Painlev\'e equation in its standard form,
but solution to the discrete Painlev\'e equations is invariant w.r.t.
multiplying the solution by this factor, see below.\label{ftn}}
\be
\label{ZB5}
Z_{N_1,N_2}^{(5d)}=z^{2\alpha_1\alpha_2}(z;q)_{2\alpha_2\alpha_3}
\cdot{1\over N_1!N_2!}
\int\prod_{i=1}^{N_1} \Big(z^{2\alpha_1+2\alpha_2+N_1}d_qu_iu_i^{2\alpha_1}(u_i;q)_{2\alpha_2}(zu_i;q)_{2\alpha_3}\Big)\Delta^2(u) \times\\ \times
\int\prod_{j=1}^{N_2} \Big(d_qv_j v_j^{-2\alpha_1-2\alpha_2-2\alpha_3-2N_1-2}(zv_j;q)_{2\alpha_2}(v_j;q)_{2\alpha_3}\Big)\Delta^2(v)\times \prod_{i=1}^{N_1}\prod_{j=1}^{N_2}\Big(1-zu_iv_j\Big)^2
\ee
where the numbers of integration $N_1$ and $N_2$ are given by the same formula (\ref{N}).

The function $Z^{(5d)}_{N_1,N_2}$ is related to the 5d Nekrasov functions $Z_{\lambda,\mu}$ via
\be
Z^{(5d)}_{N_1,N_2}=\Big(\mathfrak{Z}^{(5d)}\Big)^{-1}\cdot z^{\Delta-\Delta_1-\Delta_2 }\cdot\sum_{\lambda,\mu}\Big(q^{2\alpha_3+1}z\Big)^{|\lambda|+|\mu|}Z_{\lambda,\mu}
\ee

\subsection{The Fourier transform of the conformal block}

Now we again introduce a generating function of $Z_{N_1,N_2}^{(5d)}$, which is the Fourier transform of the $q$-Virasoro conformal block,
\be
Z_N^{(5d)}(\mu_1,\mu_2)=\sum_{N_1,N_2:\ {N_1+N_2=N}}
\mu_1^{N_1}\mu_2^{N_2}\cdot Z_{N_1,N_2}^{(5d)}
\label{FT5}
\ee
Similarly to $Z_N^{(4d)}(\mu_1,\mu_2)$, this function is a Toda chain $\tau$-function (in Miwa variables) and has a determinant representation
\be
Z_N^{(5d)}(\mu_1,\mu_2)= z^{2\alpha_1\alpha_2}(z;q)_{2\alpha_2\alpha_3}
\cdot\det_{1\le i,j\le N}G(i+j-2)
\label{detrep5}
\ee
where
\be\label{G5}
G(k)= \mu_1z^{2\alpha_{12}+k+1} \int d_qu u^{2\alpha_1+k}(u;q)_{2\alpha_2}(zu;q)_{2\alpha_3}+\mu_2
\int d_qv v^{-2\alpha_1-2\alpha_2-2\alpha_3-2-k}(zv;q)_{2\alpha_2}(v;q)_{2\alpha_3} =
\nn\\
=\mu_1\cdot z^{2\alpha_{12}+k+1}\cdot \mathfrak{B}_q(2\alpha_1+k+1,2\alpha_2+1)
\phantom{A}_2\phi_1(q^{-2\alpha_3},q^{2\alpha_1+k+1};q^{2\alpha_{12}+k+2};q,z)+\nn\\
+\mu_2\cdot q^{-(2\alpha_1+1)(2\alpha_{23}+1)}\cdot \mathfrak{B}_q(-2\alpha_{123}-k-1,2\alpha_3+1)
\phantom{A}_2\phi_1(q^{-2\alpha_{123}-k-1},q^{-2\alpha_2};q^{-2\alpha_{12}-k};q,z)
\ee
where we denote $\alpha_{12}=\alpha_1+\alpha_2$ etc,
and $\mathfrak{B}_q(\alpha,\beta)= \int_0^1 d_qxx^{\alpha-1}(x;q)_{_{\beta-1}}=
{\Gamma_q (\alpha)\Gamma_q (\beta)\over\Gamma_q (\alpha+\beta)}$
is the $q$-Beta-function constructed from the $q$-$\Gamma$-functions \cite{GasR},
while $\phantom{A}_2\phi_1(a,b;c;q,z)$
is the Heine basic $q$-hypergeometric function \cite{GasR},
\be
\phantom{A}_2\phi_1(a,b;c;q,z):=\sum_{n=0}^\infty {(a;q)_n(b;q)_n\over (c;q)_n(q;q)_n}\ z^n
\ee

The determinant representation, similarly to $Z_N^{(4d)}(\mu_1,\mu_2)$, follows from the eigenvalue representation
\be\label{qMM}
Z_N^{(5d)}(\mu_1,\mu_2)\sim{1\over N!}
\int\prod_{i=1}^{N} \Big(d_qx_ix_i^{2\alpha_1}(z^{-1}x_i;q)_{2\alpha_2}(x_i;q)_{2\alpha_3}\Big)\Delta^2(x)
\ee

\subsection{Conformal block as a discrete Painlev\'e solution}

We define now the function\footnote{Note the notation here differs from that in \cite{Jimbo17}, where the correspondence between the $q$-conformal block and the discrete Painlev\'e VI equations was established for a non-matrix model case, when (\ref{N}) has not to be satisfied.} $\tau(\alpha_1,\alpha_2,\alpha_3,\alpha_4;z)=\tau_N(\alpha_i;\mu_1/\mu_2,z)
= \mu_2^NZ_N^{(5d)}(\mu_1,\mu_2)z^{-2\alpha_1\alpha_2}(z;q)^{-1}_{2\alpha_2\alpha_3}$, for simplicity of notation removing the simple factor $z^{2\alpha_1\alpha_2}(z;q)_{2\alpha_2\alpha_3}$ (as we noted above, see footnote \ref{ftn}, multiplying the $\tau$-function with the factor $(z;q)_{2\alpha_2\alpha_3}$ does not change the solution to the Painlev\'e equation) and using that $N=-\alpha_1-\alpha_2-\alpha_3-\alpha_4$. Then, we have, in fact, eight different $\tau$-functions that  are used for constructing the discrete Painlev\'e equations:
\be\label{tau}
\tau_1(\alpha_1,\alpha_2,\alpha_3,\alpha_4;z)=\tau(\alpha_1+{1\over 2},\alpha_2,\alpha_3+{1\over 2},\alpha_4;z)& \ \ \ \ \ \
\tau_2(\alpha_1,\alpha_2,\alpha_3,\alpha_4;z)=\tau(\alpha_1,\alpha_2-{1\over 2},\alpha_3,\alpha_4+{1\over 2};z)&\nn\\ \tau_3(\alpha_1,\alpha_2,\alpha_3,\alpha_4;z)=\tau(\alpha_1,\alpha_2,\alpha_3+{1\over 2},\alpha_4+{1\over 2};z)& \ \ \ \ \ \ \tau_4(\alpha_1,\alpha_2,\alpha_3,\alpha_4;z)=\tau(\alpha_1+{1\over 2},\alpha_2-{1\over 2},\alpha_3,\alpha_4;z)\nn\\
\tau_5(\alpha_1,\alpha_2,\alpha_3,\alpha_4;z)=\tau(\alpha_1+{1\over 2},\alpha_2,\alpha_3,\alpha_4+{1\over 2};z)&\ \ \ \ \ \ \tau_6(\alpha_1,\alpha_2,\alpha_3,\alpha_4;z)=\tau(\alpha_1,\alpha_2-{1\over 2},\alpha_3+{1\over 2},\alpha_4;z)\nn\\ \tau_7(\alpha_1,\alpha_2,\alpha_3,\alpha_4;z)=\tau(\alpha_1+{1\over 2},\alpha_2-{1\over 2},\alpha_3+{1\over 2},\alpha_4+{1\over 2};z)&\ \ \ \ \ \ \tau_8(\alpha_1,\alpha_2,\alpha_3,\alpha_4;z)=\tau(\alpha_1,\alpha_2,\alpha_3,\alpha_4;z)
\ee
Indeed, one can construct the functions $w_i(z)$ through these 8 different $\tau$-functions in accordance with the weight lattice of $D_5^{(1)}$ \cite{Sakai,TM,Jimbo17} (in fact, due to bilinear relations \cite{Jimbo17}, they can be expressed through 4 $\tau$-functions):
\be\label{wtau}
w_1(z)=q^{N}z\cdot{\tau_1(qz)\tau_2(z)\over\tau_3(qz)\tau_4(z)}\nn\\
w_2(z)=q^{2\alpha_3+2N-1}z\cdot{\tau_5(z)\tau_6(z)\over \tau_7(z)\tau_8(z)}
\ee
and these functions $w_1(z)$ and $w_2(z)$ satisfy the discrete Painlev\'e $q$-PVI equations (\ref{PVI}) with
\be\label{param}
-N=\sum_i\alpha_i,\ \ \ \ \ \ \ a_1=q,\ \ \ \ a_2=q^{1-N-2\alpha_3},\ \ \ \ a_3=q^{2-N},\ \ \ \ a_4=q^{2\alpha_2+1}\nn\\
b_1=q^{-2\alpha_2+1},\ \ \ \ b_2=q^{2\alpha_1+2\alpha_3+N+1},\ \ \ \ b_3=q^{2\alpha_3+1},\ \ \ \ b_4=q^{2\alpha_1+2\alpha_3+N+1}
\ee
The first constraint in this list allows us to omit $\alpha_4$  from the set of the arguments
of $\tau$-functions (\ref{tau}).
As we noted earlier, one can always express $a_i$, $b_i$ through any four independent parameters, four $\alpha_i$ in this case. A determinant solution to the $q$-Painlev\'e equation was also obtained in \cite{Sakai}. After some manipulations with the $q$-hypergeometric functions, it can be reduced to solution (\ref{detrep5}) at $\mu_2=0$.

\section{Discrete Painlev\'e: integrability in Miwa variables + string equations}

\subsection{Conformal matrix model and Hirota bilinear identities}

One can check that these $\tau$-functions satisfy the eight bilinear relations \cite{Sakai,Jimbo17}:
\be
zq^{2N-2}\tau_1\tau_2-q^{2\alpha_2}\tau_3\tau_4-\tau_7\tau_8=0
\label{bi1}
\ee
\be
\tau_1\tau_2-q^{1-2\alpha_3-2N}\tau_3\tau_4-\tau_5\tau_6=0
\label{bi2}
\ee
\be
\overline{\tau}_1\tau_2-q^{1-N}\overline{\tau}_3\tau_4-q^{2N-2\alpha_{12}-2}\tau_5\overline{\tau}_6=0
\label{bi4}\\
\nn\\
zq^{N-1}\overline{\tau}_1\tau_2-q^{2\alpha_2}\overline{\tau}_3\tau_4-\overline{\tau}_7\tau_8=0
\label{bi5}\\
\nn\\
z\overline{\tau}_1\tau_2-q^{2-2N}\overline{\tau}_3\tau_4-q^{-2\alpha_2}\tau_7\overline{\tau}_8=0
\label{bi3}
\ee
\be
\overline{\tau}_1\tau_2-q^{1-2N-2\alpha_3}\overline{\tau}_3\tau_4-\overline{\tau}_5\tau_6=0
\label{bi6}
\ee
\be
\overline{\tau}_1\underline{\tau}_2-q^{1-N}\overline{\tau}_3\underline{\tau}_4-q^{2N-2-2\alpha_{12}}\tau_5\tau_6=0
\label{bi7}
\ee
\be
z\overline{\tau}_1\underline{\tau}_2-q^{2-N}\overline{\tau}_3\underline{\tau}_4-q^{N-2\alpha_{2}}\tau_7\tau_8=0
\label{bi8}
\ee
where we introduced the standard notation $\overline{\tau}:=\tau(qz)$, $\underline{\tau}:=\tau(q^{-1}z)$. From these identities, one can derive the discrete Painlev\'e $q$-PVI equations \cite{Sakai}.

These bilinear identities can be derived in many various ways, important for us is that the first four of these bilinear identities can be obtained from the matrix model representation (\ref{qMM}) exploiting the fact that this latter is a $\tau$-function of the Toda chain hierarchy \cite{Toda}. Indeed, note that the multiple integral (\ref{qMM}) can be also presented in the form (\ref{Z4t}), with integral substituted by the Jackson integral (which is inessential for integrable properties and, hence for the Hirota identities) and three sets of Miwa variables
\be\label{rMiwa}
(0,2\alpha_1),\ \ \ (zq^{-i},1),\ \ i=0,\ldots,2\alpha_2-1,\ \ \ (q^{-i},1),\ \ i=0,\ldots,2\alpha_3-1
\ee
for integer $2\alpha_2$ and $2\alpha_3$. At the same time, this eigenvalue integral (\ref{qMM}) is not only a $\tau$-function of the Toda chain, it is simultaneously a $\tau$-function of the discrete Toda chain \cite{MMV}.

As a $\tau$-function, the integral (\ref{qMM}) also satisfies the Hirota identities (\ref{Hirota1})-(\ref{Hirota4}). For instance, choosing $(z_a,\alpha_a)=(0,\alpha_1)$, $(z_b,\alpha_b)=(q^{-2\alpha_3+1},1)$ and $(z_c,\alpha_c)=(q^{-2\alpha_2+1},1)$ and using (\ref{tau}), one obtains from (\ref{Hirota3}) the bilinear identity (\ref{bi2}). Similarly, choosing $(z_a,\alpha_a)=(0,\alpha_1)$, $(z_b,\alpha_b)=(q^{-2\alpha_3},1)$ and $(z_c,\alpha_c)=(q^{-2\alpha_2+2},1)$, one obtains from (\ref{Hirota3}) the bilinear identity (\ref{bi1}). In order to obtain these formulas, one has to take into account a normalization factor that gives rise to additional factors like $q^{2N}$ in the coefficients of the bilinear identities.

Similarly, one can note that the rescaling $\tau(\alpha_3)\to\overline{\tau}(\alpha_3+1/2)$ corresponds to adding the Miwa variable with unit multiplicity at the point $qz$. Considering this as $z_b$ with $(z_a,\alpha_a)=(0,\alpha_1)$ and $(z_c,\alpha_c)=(q^{-2\alpha_2+1},1)$ and using (\ref{tau}), one immediately obtains (\ref{bi4}) from (\ref{Hirota3}). At last, one can obtain, in a similar way, (\ref{bi5}).

\subsection{Discrete Painlev\'e equation and the string equation}

Note that the four integrable Hirota identities are not enough to fix a solution to the discrete Painlev\'e equations, the integral (\ref{qMM}), because they are satisfied by integrals with an arbitrary measure $\mu(x)$. To put it differently, they encode just an integrable hierarchy, which has a lot of different solutions, and (\ref{qMM}) is only one of them. In order to fix this concrete solution, one needs additional constraints, and these constraints are the Virasoro constraints considered at the point with only 3 non-zero Miwa variables. This is because the Ward identities (Virasoro constraints) crucially depend on the chosen measure function $\mu(x)$. The Ward identities typically fix the solution up to a choice of integration contours. In the present case with the matrix model (\ref{qMM}), there are, at least, two solutions, which are the two $q$-hypergeometric functions in (\ref{G5}). In the non-discrete case, there is an argument that the Ward identities leave no room for more solutions: the partition function (\ref{MM}) is associated, as usual for the Dijkgraaf-Vafa solution, with two possible extrema (minima with a proper choice of parameters) of the matrix (eigenvalue) model potential: it is a sum of three logarithms that exactly has two extrema.

Thus, the four Virasoro constraints play the role of the string equation: the string equation added to integrability leads to the discrete Painlev\'e $q$-PVI equations. This is much similar to the way the usual Painlev\'e equation emerges from the string equation of the matrix models \cite{UFN3} (e.g. the Painlev\'e I equation for the Kontsevich model, see sec.\ref{Kont} and eq.(\ref{PI})).

\subsection{An illustration: $N=1$ case}

In order to illustrate the phenomenon, we consider the simplest case of $N=1$ ``matrix model" (\ref{Z4M}) with $\mu(x)=1$, i.e. the matrix of size $1\times 1$ and just one integration \cite{qPsh}. In this case,
\be\label{sv}
{b_1\over b_3}=q{a_2\over a_4}=q^{-2\alpha_{23}},\ \ \ \ {b_2\over b_4}={a_1\over a_3}=1
\ee
as follows from (\ref{param}). With these special values of parameters (\ref{sv}), the discrete Painlev\'e $q$-PVI equations (\ref{PVI}) admits solutions that satisfy a simpler pair of equations
\be\label{lPVI}
w_1(z)=a_4{w_2(z)-b_1z/q\over w_2(z)-b_3},\ \ \ \ \ \ \ w_2(qz)=b_4{w_1(z)-a_1z\over w_1(z)-a_3}
\ee
Indeed, from the second equation it follows that
\be
w_1(z)={a_3w_2(qz)-b_4a_1z\over w_2(qz)-b_4}\ \stackrel{(\ref{sv})}{=}\ a_3{w_2(qz)-b_2\over w_2(qz)-b_4}
\ee
Multiplying it with the first equation taken at $z\to qz$, one obtains the first equation from (\ref{PVI}). Similarly, it follows from the first equation that
\be
w_2(z)={b_3w_1(z)-b_1za_4/q\over w_1(z)-a_4}\ \stackrel{(\ref{sv})}{=}\ b_3{w_1(z)-a_2\over w_1(z)-a_4}
\ee
Multiplying it with the second equation, we obtain the second equation from (\ref{PVI}).

In this case, the bilinear identities become linear and (\ref{bi5}) follows from (\ref{bi1}), (\ref{bi6}) from (\ref{bi2}), (\ref{bi7}) from (\ref{bi4}) and (\ref{bi8}) from (\ref{bi3}) so that the independent four identities are
\be
z\tau_2-{q\over b_1}\tau_4-\tau_8=0
\label{li1}\\
\nn\\
\tau_2-{1\over b_3}\tau_4-\tau_6=0
\label{li2}\\
\nn\\
\tau_2-\tau_4-{qb_1\over a_2b_2}\overline{\tau}_6=0
\label{li3}\\
\nn\\
z\tau_2-\tau_4-{b_1\over q}\overline{\tau}_8=0
\label{li4}
\ee
Now expressing $\tau_6$ and $\tau_8$ from (\ref{li1}) and (\ref{li2}) and substituting them into (\ref{wtau}), we obtain that the first equation of (\ref{lPVI}) is, indeed, correct. Similarly, expressing $\overline{\tau}_6$ and $\overline{\tau}_8$ from (\ref{li3}) and (\ref{li4}), we prove the second equation of (\ref{lPVI}).

Of these four identities, only the last one (\ref{li4}) is not a corollary of integrability and is correct only for the specific $\mu(x)=1$ measure. Hence, it should be just the string equation. Let us analyze it in detail.

Note that, at $N=1$, the first string equation in Miwa variables reads
\be
\sum_a 2\alpha_aZ_1(z_a;\alpha_a-1/2)=0
\ee
where the sum goes over all Miwa variables.
In the case of a restricted number of Miwa variables (\ref{rMiwa}), when (\ref{Z4M}) reduces to (\ref{qMM}) at $N=1$, this equation turns into
\be\label{1}
\hat L_{-1}\tau = 0 &\Longrightarrow &
[2\alpha_1]_q\cdot\tau(\alpha_1-1/2)-[2\alpha_2]_q\cdot\tau(\alpha_2-1/2)-q^{2\alpha_1-2\alpha_3}
[2\alpha_3]_q\cdot\tau(\alpha_3-1/2)=0
\ee
Here $[n]_q:=(1-q^n)/(1-q)$ denotes the quantum numbers.
However, this lowest $L_{-1}$ constraint does not contain $z$ and  is not just the same as (\ref{li4}).
Fortunately,
there are more equations, those associated with  $\hat L_1$ and $\hat L_2$ Virasoro constraints
(all other Borel Virasoro generators are obtained by repeated commutation of $\hat L_2$ and $\hat L_{\pm 1}$), and they involve $z$.
Thus we can try (and succeed) to get (\ref{li4}) by adding these constraints to (\ref{1}).
In fact, $\hat L_2$ is needed
when one deals with an arbitrary number of Miwa variables.
In the $N=1$ case and the number of Miwa variables restricted to the set (\ref{rMiwa}),
one can substitute it by a much simpler $\hat L_0$.
In this case, the two independent constraints in addition to $\hat L_{-1}$ are  \cite{qPsh}\footnote{Note that the definitions in that paper are slightly different,
thus the formulas are slightly different as well .}:
\be
\begin{array}{ccl}
\hat L_0\tau=0 &\Longrightarrow &[-2\alpha_{1}]_q(q-q^{-2\alpha_3}z)\cdot\tau(\alpha_1-1/2)+
q^{2\alpha_2}[2\alpha_2]_q(q-z)\tau(\alpha_2-1/2,z/q)-\\
&&\\
&&-[-2\alpha_3](q\cdot
q^{2\alpha_2}-q^{-2\alpha_3}z)\tau(\alpha_3-1/2)=0\\
&&\\
\hat L_1\tau=0 &\Longrightarrow &
\tau+q^{2\alpha_2}\cdot\tau(\alpha_1+1/2,\alpha_2-1/2)-z\cdot\tau(\alpha_2-1/2)=0
\end{array}
\ee
In fact, the second equation is a combination of $\hat L_1$-constraint and the lower ones, and is nothing but equation (\ref{li1}).
Now, one can obtain from the first two Virasoro constraints that
\be
\left\{
\begin{array}{lr}
\phantom{.}[2\alpha_1]_q(q-1)\tau(\alpha_1-1/2)=q^{-2\alpha_3-1}(q^{2\alpha_{23}+1}-z)\tau(\alpha_2-1/2)
-q^{2\alpha_{12}-1}(q-z)\underline{\tau}(\alpha_2-1/2)&\\
&\Longrightarrow\\
\phantom{.}[2\alpha_3]_q(q-1)\tau(\alpha_3-1/2)=
q^{-2\alpha_1-1}(q^{2\alpha_{3}+1}-z)\tau(\alpha_2-1/2)-q^{2\alpha_{23}-1}(q-z)
\underline{\tau}(\alpha_2-1/2)&
\end{array}
\right.
\label{159}
\ee
Replacing in the first of these equations $\alpha_1\to\alpha_1+1/2$ and, in the second, $\alpha_3\to\alpha_3+1/2$, one finally obtains from the first two Virasoro constraints the identities
\be\label{lai1}
\Big({a_2b_2\over q}-1\Big)\overline{\tau}_8=\Big({q\over b_1}-a_2z\Big)\overline{\tau}_4
-{a_2b_2\over b_1}(1-z){\tau}_4\\
(b_3-1)\overline{\tau}_8=
{q^2\over a_2b_2}(b_3-z)\overline{\tau}_6-{q^2\over b_1a_2}(1-z)
\tau_6
\label{lai2}
\ee
and the second of these identities is exactly (\ref{li4}) provided one applies (\ref{li2}) and (\ref{li3}) to express $\tau_6$ and $\overline{\tau}_6$ through $\tau_2$ and $\tau_4$.

\section{Conclusion
\label{conc}}

In this letter, we made and justified the following set of statements:
\begin{itemize}
\item The Fourier transform of the $q$-conformal block has a manifest determinant representation when is presented by the conformal matrix model.
\item This determinant solves the discrete Painlev\'e $q$-PVI equations.
\item This discrete Painlev\'e solution follows from a combination of integrability and string equations of the matrix model in Miwa variables restricted to a particular set of Miwa variables.
\end{itemize}
For the sake of illustration, we considered the $N=1$ case in a very detail. More technical issues are postponed to an expanded version of this text.

The main message of this paper is that the string equations can be significantly less naive
than just the lowest Virasoro constraints.
Moreover, it calls for a deeper understanding of the structure and the shape of Ward identities
in logarithmic models and in Miwa variables, which are getting more and more important in
modern theory.
If the first emergency of the simplest equations from the Painl\'eve family in the double scaling
limit of Hermitian model \cite{dsH} was long considered to be just an accident,
our work demonstrates that things are very different: the Painl\'eve equations seem to appear
{\it naturally} within this context, and, if so, one needs to understand what has the
Painl\'eve property to do with the Virasoro constraints.
This adds to the long-standing puzzle of the Painl\'eve property of reductions of integrable systems to ODE \cite{PT}.
Last, but not the least, we once again confirmed the relative simplicity of the
q-Painl\'eve equations as compared to the continuous ones,
and this emphasizes the  importance of clearly defining the Painleve property of finite-difference
equations and of clearly describing the limiting procedure connecting it to the continuous case.
As we explained, this involves study of the condensation of Miwa variables and the related problem
of various phase transitions in the space of $\tau$-functions.
Hopefully the  identity\\

\centerline{
{\bf Painl\'eve = string }
}

\bigskip

\noindent
in the space of difference/differential equations
looks impressive  and challenging enough
to give a new momentum for work in all these directions.

\section*{Acknowledgements}

Our work is supported in part by the grant of the Foundation for the Advancement of Theoretical Physics ``BASIS" (A.Mir., A.Mor.), by  RFBR grants 19-01-00680 (A.Mir.,Z.Z.) and 19-02-00815 (A.Mor.), by joint grants 19-51-53014-GFEN-a (A.Mir., A.Mor.), 19-51-50008-YaF-a (A.Mir.), 18-51-05015-Arm-a (A.Mir., A.Mor.), 18-51-45010-IND-a (A.Mir., A.Mor.). The work was also partly funded by RFBR and NSFB according to the research project 19-51-18006 (A.Mir., A.Mor.).

\end{document}